\documentclass[reprint,preprintnumbers,prd,nofootinbib]{revtex4-1}

\usepackage{amsfonts}
\usepackage{amsmath}
\usepackage{array}
\usepackage{braket}
\usepackage{epstopdf}
\usepackage{graphicx}
\usepackage{hepparticles}
\usepackage{hepnicenames}
\usepackage{hepunits}
\usepackage{hyperref}
\usepackage[%
    utf8
]{inputenc}
\usepackage{slashed}
\usepackage{subfigure}
\usepackage{placeins}
\usepackage[%
    normalem
]{ulem}
\usepackage[%
    usenames,
    svgnames,
    dvipsnames
]{xcolor}

\newcommand{\para}{\parallel}

\newcommand{\dd}{\ensuremath{\textrm{d}}}

\newcommand{\eqa}[1]{\begin{eqnarray} #1 \end{eqnarray}}

\def\qsq       {\ensuremath{q^2}\xspace}

\def\ellell     {\ensuremath{\ell^+ \ell^-}\xspace}
\def\mumu       {{\ensuremath{\Pmu^+\Pmu^-}}\xspace}

\def\lhcb {\mbox{LHCb}\xspace}
\def\belle  {\mbox{Belle}\xspace}

\def\WC  {\ensuremath{\mathcal{C}}\xspace}

\begin{document}

\allowdisplaybreaks

\preprint{ZU-TH-15/18}
\title{Towards establishing lepton flavour universality violation in $\bar{B}\to \bar{K}^*\ell^+\ell^-$ decays}
\author{Andrea Mauri}
\email{a.mauri@cern.ch}
\author{Nicola Serra}
\email{nicola.serra@cern.ch}
\author{Rafael Silva Coutinho}
\email{rafael.silva.coutinho@cern.ch}
\affiliation{Physik-Institut, Universit\"at Z\"urich, Winterthurer Strasse 190, 8057 Z\"urich, Switzerland}

\begin{abstract}
  Rare semileptonic $b \to s \ell^+ \ell^-$ transitions provide some of the most promising frameworks to search for new physics effects. 
Recent analyses of these decays have indicated an anomalous behaviour in measurements
  of angular distributions of the decay $B^0\to
    K^*\mu^+\mu^-$ and lepton-flavour-universality observables. 
Unambiguously establishing if these  deviations have a common nature
 is of paramount importance in order to understand the observed pattern. 
  We propose a novel approach to independently and complementary
  probe this hypothesis   
  by performing a simultaneous amplitude analysis of $\bar{B}^0 \to \bar{K}^{*0} \mu^+\mu^-$  and $\bar{B}^0 \to \bar{K}^{*0} e^+e^-$ decays. 
  This method enables the direct determination of  
  observables that encode potential non-equal couplings of muons and electrons,  
  and are found to be insensitive to nonperturbative QCD effects. 
If current hints of new physics are confirmed, our approach could
allow an early discovery of physics beyond the standard
model with LHCb run II data sets. 
\end{abstract}

\maketitle

Flavour changing neutral current processes of {\textit{B}} meson decays 
are crucial probes for the standard model (SM), 
since as yet undiscovered particles may contribute to these transitions and cause observables to deviate  
from their SM predictions~\cite{Grossman:1996ke,Fleischer:1996bv,London:1997zk,Ciuchini:1997zp}.
The decay mode $\bar{B}\to \bar{K}^*\ell^+\ell^-$ is a prime example (i.e., $\ell = \mu, e$), 
which offers a rich framework to study from differential decay widths to angular observables.
An anomalous behaviour in angular and branching fraction analyses of the decay channel 
$\bar{B}^{0} \to \bar{K}^{*0} \mu^{+}\mu^{-}$ has been recently reported~\cite{Aaij:2015oid,Wehle:2016yoi,Aaij:2013aln,Aaij:2014pli}, 
notably in one of the observables with reduced theoretical uncertainties, 
$P^{\prime}_{5}$~\cite{Aaij:2013qta,Descotes-Genon:2015uva}.
Several models have been suggested in order to interpret these results as new physics (NP) 
signatures~\cite{Gauld:2013qja,Buras:2013qja,Altmannshofer:2013foa,Crivellin:2015era,Hiller:2014yaa,Biswas:2014gga,Gripaios:2014tna}.  
Nonetheless, the vectorlike nature of this pattern could be also explained by 
non-perturbative QCD contributions from $b\to s c{\bar{c}}$ operators (i.e., charm loops) 
that are able to either mimic or camouflage NP effects~\cite{Jager:2012uw,Jager:2014rwa,Ciuchini:2015qxb}. 
Nonstandard measurement in ratios of $b \to s \ell^+ \ell^-$ processes 
- such as of $R_{K}$~\cite{Aaij:2014ora} and $R_{K^{*}}$~\cite{Aaij:2017vbb} - 
indicate a suppression of the muon channel which is also compatible with the $P^{\prime}_{5}$ anomaly. 
In this case an immediate interpretation of lepton flavour universality (LFU) breaking is 
suggested due to the small theoretical uncertainties in their predictions~\cite{Hiller:2003js,Bordone:2016gaq}. 
Whilst the individual level of significance of the present anomalies is still inconclusive, 
there is an appealing nontrivial consistent pattern shown in 
global analysis fits~\cite{Capdevila:2017bsm,Altmannshofer:2017yso,Hurth:2017hxg,Ciuchini:2017mik,Alok:2017sui}.

The formalism of {\textit{b}} decays is commonly described within 
an effective field theory~\cite{Altmannshofer:2008dz}, 
which probes distinct energy scales; 
with regimes classified into short-distance (high energies) perturbative 
and noncalculable long-distance effects. 
These can be parametrised in the weak Lagrangian in terms of 
effective operators with different Lorentz structures, $\mathcal{O}_i$, 
with corresponding couplings $\mathcal{C}_i$ - referred to as Wilson coefficients (WC). 
Only a subset of the operators that are most sensitive to NP is examined in this work~\cite{Bobeth:2017vxj}, 
\textit{i.e.} $\mathcal{O}_7$ (virtual photon exchanges), $\mathcal{O}_{9,10}$ (vector and axial currents) 
and corresponding right-handed couplings with flipped helicities. 
In this framework NP effects are incorporated 
by introducing deviations in the WCs~\cite{Ali:1994bf} from their SM predictions,  
i.e., $\mathcal{C}_i = \mathcal{C}^{\mathrm{SM}}_i + \mathcal{C}^{\mathrm{NP}}_i$. 
For instance, the anomalous pattern seen in semileptonic decays can be 
explained by a shift in the coefficient $\mathcal{C}_9$ only, 
or $\mathcal{C}_9$ and $\mathcal{C}_{10}$ simultaneously~\cite{Capdevila:2017bsm,Altmannshofer:2017yso,Hurth:2017hxg}. 
A direct experimental determination of the WCs is currently 
bounded by sizeable uncertainties that arise from 
nonfactorisable hadronic matrix elements that are difficult to assess reliably from first principles. 
Some promising approaches suggest to extract this contribution 
from data-driven analyses~\cite{Blake:2017fyh,Hurth:2017sqw} 
and by exploiting analytical properties of its structure~\cite{Bobeth:2017vxj}. 
However, these models still have intrinsic limitations, in particular 
in the assumptions that enter in parametrisation of the dilepton invariant mass distribution. 

In this article, we propose a new model-independent approach that 
from a simultaneous unbinned amplitude analysis of both  
$\bar{B}^0 \to \bar{K}^{*0} \mu^+\mu^-$  and $\bar{B}^0 \to \bar{K}^{*0} e^+e^-$ decays 
can, for the first time, unambiguously determine LFU-breaking from direct measurements of WCs. 
This work builds on the generalisation of Ref.~\cite{Bobeth:2017vxj}, 
but it is insensitive to the model assumptions of the parametrisation. 
This effect relies on the strong correlation between the muon and electron modes 
imposed by the lepton-flavour universality of the hadronic matrix elements. 
Furthermore, in this method the full set of observables  
(\textit{e.g} $R_{K^{*}}$, $P^{\prime}_{5}$ and branching fraction measurements) available in $\bar{B}\to \bar{K}^*\ell^+\ell^-$  
decays is exploited, providing unprecedented precision on LFU in a single analysis.

Consider the differential decay rate for $\bar{B}\to \bar{K}^*\ell^+\ell^-$ 
decays (dominated by the on-shell $\bar{K}^{*0}$ contribution) 
fully described by four kinematic variables;  
the di-lepton squared invariant mass, $q^2$, and the three angles 
$\vec{\Omega} = (\cos \theta_\ell, \cos \theta_K, \phi)$~\cite{Altmannshofer:2008dz}.
The probability density function ($p.d.f.$) for this decay can be written as
\begin{equation}
p.d.f.  = \frac{1}{\Gamma} \frac{\dd^4 \Gamma}{\dd q^2 \dd^3 \Omega}, \
    \quad
    \text{with}\quad
    \Gamma = \int_{q^2} \dd q^2 \frac{\dd\Gamma}{\dd q^2}\,,
\end{equation}
with different \qsq intervals depending on the lepton flavour under study. 
For a complete definition of $\dd ^4\Gamma/(\dd q^2 \dd ^3\Omega)$ we refer 
to~\cite{Bobeth:2008ij,Altmannshofer:2008dz} and references therein.
It is convenient to explicitly write the WC dependence on the decay width by 
the transversity amplitudes ($\lambda=\perp, \para,0$) as~\cite{Bobeth:2017vxj}
\eqa{
  \label{eq:amp_dep}
  {\cal{A}}_{\lambda}^{(\ell)\,L,R} &=& {\cal{N}}_{\lambda}^{(\ell)}\ \bigg\{ 
(C^{(\ell)}_9 \mp C^{(\ell)}_{10}) {\cal{F}}_{\lambda}(q^2) \\
&&+\frac{2m_b M_B}{q^2} \bigg[ C^{(\ell)}_7 {\cal{F}}_{\lambda}^{T}(q^2) - 16\pi^2 \frac{M_B}{m_b} {\cal{H}}_{\lambda}(q^2) \bigg]
\bigg\}\,,\nonumber 
}
where ${\cal{N}}_{\lambda}^{(\ell)}$ is a normalisation factor, and 
${\cal{F}}^{(T)}_{\lambda}(q^{2})$ and $\mathcal{H}_\lambda(q^{2})$ 
are referred to ``local'' and ``nonlocal'' hadronic matrix elements, respectively.  
The ${\cal{F}}^{(T)}_{\lambda}(q^{2})$ are form factors, 
while $\mathcal{H}_\lambda(q^{2})$ encode the aforementioned nonfactorisable
hadronic contributions and
are described using two complementary parametrisations~\cite{Bobeth:2017vxj,Hurth:2017sqw} - 
for brevity only a subset of results is shown for the latter approach.
In the following
this function is expressed in terms of a ``conformal'' 
variable $z(q^{2})$~\cite{Bobeth:2017vxj,Boyd:1995cf,Bourrely:2008za}, 
with an analytical expansion truncated at a given order  
$z^n$ (herein referred to as $\mathcal{H}_\lambda[z^n]$), 
after removing singularities related to the $J/\psi(1S)$ and $\psi(2S)$. 
Further information about the formalism is given in appendix~\ref{app:formalism}.
One of the drawbacks of this expansion is that \textit{a priori} there is 
no physics argument to justify the order of the polynomial to be curtailed at - 
which in turn currently limits any claim on NP sensitivity. 

In order to overcome these points, we investigate the LFU-breaking
hypothesis using direct determinations of the difference of Wilson coefficients 
between muons and electrons, i.e.,
\begin{equation}
\Delta \WC_i = \widetilde{\mathcal{C}}_i^{(\mu)} - \widetilde{\mathcal{C}}_i^{(e)}\,,
\end{equation}
where the usual WCs $\mathcal{C}_i^{(\mu,e)}$ are renamed as $\widetilde{\mathcal{C}}_i^{(\mu,e)}$, 
since an accurate disentanglement between the physical 
meaning of $\WC_i^{(\mu,e)}$ and the above-mentioned hadronic pollution 
cannot be achieved at the current stage of the theory~\cite{Chrzaszcz:2018yza}.
The key feature of this strategy is to realise that all hadronic matrix elements 
are known to be lepton-flavour universal, and thus are shared among both semileptonic decays.
This benefits from the large statistics available for $\bar{B}^0 \to \bar{K}^{*0} \mu^+\mu^-$ decays 
that is sufficient to enable the determination of these multispace parameters. 
Note that an amplitude analysis of the electron mode only has been previously disregarded, 
given the limited data set in either LHCb or Belle experiments.
In a common framework the hadronic contributions are treated as 
nuisance parameters, while only the Wilson coefficients $\widetilde{\WC}_9^{(\mu,e)}$ 
and $\widetilde{\WC}_{10}^{(\mu,e)}$ are kept separately for the two channels.
For consistency the WC $\widetilde{C}_{7}$ is also shared in the fit 
and fixed to its SM value, given its universal coupling to photons
and the strong constraint from radiative $B$ decays~\cite{Paul:2016urs}. 
In the following, all the right-handed WCs are fixed to their SM values, i.e., $\WC_i^{\prime\,(\mu,e)} = 0$,
while sensitivity studies on the determination of the WCs $\WC_9^{\prime\,(\mu)}$ and  $\WC_{10}^{\prime\,(\mu)}$
are detailed in Appendix~\ref{app:WC_primed}.

Signal-only ensembles of pseudoexperiments are generated with
sample size corresponding roughly to the yields foreseen in LHCb run-II [$8\,$fb$^{-1}$] and future upgrades 
[$50\,$-$\,300\,$fb$^{-1}$]~\cite{Aaij:2244311}, and Belle II [$50\,$ab$^{-1}$].
These are extrapolated from Refs.~\cite{Aaij:2015oid,Aaij:2017vbb,Wehle:2016yoi} 
by scaling, respectively, with luminosity and $\sigma_{b\bar{b}} \propto \sqrt{s}$ 
for LHCb, where $s$ denotes the designed centre-of-mass energy of the $b$-quark pair, 
and exclusively with luminosity for Belle II.
Note that for brevity most of the results are shown for the representative
scenario of LHCb run II. 
The studied \qsq range corresponds to 
$1.1\,\GeV^2 \leq q^2 \leq 8.0\,\GeV^2$ and $11.0\,\GeV^2  \leq q^2 \leq 12.5\,\GeV^2$ for 
the muon mode and $1.1\,\GeV^2  \leq q^2 \leq 7.0\,\GeV^2$ for the electron mode in LHCb;
while in Belle II the same kinematic regions are considered for both semileptonic channels, namely 
$1.1\,\GeV^2  \leq q^2 \leq 8.0\,\GeV^2$ and $10.0\,\GeV^2  
\leq q^2 \leq 13.0\,\GeV^2$. 
This definition of \qsq ranges are broadly consistent with published results, 
and assumes improvements in the electron mode resolution for LHCb~\cite{Lionetto:2624938}.  

Within the SM setup the Wilson coefficients are set to  
$\mathcal{C}^{\rm{SM}}_9 = 4.27$, $\mathcal{C}^{\rm{SM}}_{10} = - 4.17$ and $\mathcal{C}^{\rm{SM}}_7 = -0.34$ (see~\cite{Bobeth:2017vxj} and references therein), 
corresponding to a fixed renormalisation scale of $\mu = 4.2\,$GeV.
This baseline model is modified for two NP benchmark points (BMP), 
$\Delta\WC_9 = - 1$ and $\Delta\WC_9 = - \Delta\WC_{10} = - 0.7$,
referred to, respectively, as \texttt{BMP}$_{\WC_9}$ and \texttt{BMP}$_{\WC_{9,10}}$,  
where NP is inserted only in the case of muons, i.e., $\WC_i^{(e)} = \WC_i^{\rm{SM}}$.
These points are favoured by several global fit 
analyses with similar significance~\cite{Capdevila:2017bsm,Altmannshofer:2017yso,Hurth:2017hxg}. 

An extended unbinned maximum likelihood fit is performed to these simulated samples, 
in which multivariate Gaussian terms are added to the log-likelihood to incorporate prior knowledge
on the nuisance parameters.
In order to probe the model-independence of the framework, the nonlocal hadronic 
parametrisation is modified in several ways 
(see Appendix~\ref{app:formalism} for a detailed discussion), i.e.,
\begin{enumerate}
    \item[i.] baseline $\mathcal{H}_\lambda[z^2]$ SM prediction~\cite{Bobeth:2017vxj} 
    included as a multivariate Gaussian constraint;
    \item[ii.] no theoretical assumption on $\mathcal{H}_\lambda[z^2]$ 
    and with free-floating parameters;
  \item[iii.] higher orders of the analytical expansion of $\mathcal{H}_\lambda[z^{n}]$ 
    up to $z^3$ and $z^4$ - free floating;
    \item[iv.] and re-parametrisation of the nonlocal hadronic matrix elements as  
      proposed in Ref.~\cite{Hurth:2017sqw}, i.e.,
	including them as multiplicative factors to the corresponding leading hadronic terms.
\end{enumerate}
On the other hand, form factors parameters are taken from~\cite{Straub:2015ica} and, in order 
to guarantee a good agreement between Light-Cone Sum Rules~\cite{Ball:1998kk,Khodjamirian:2006st} 
and Lattice results~\cite{Becirevic:2006nm,Horgan:2013hoa}, 
their uncertainties are doubled with respect to Ref.~\cite{Straub:2015ica}.

\begin{figure}[t]
\includegraphics[width=.4\textwidth]{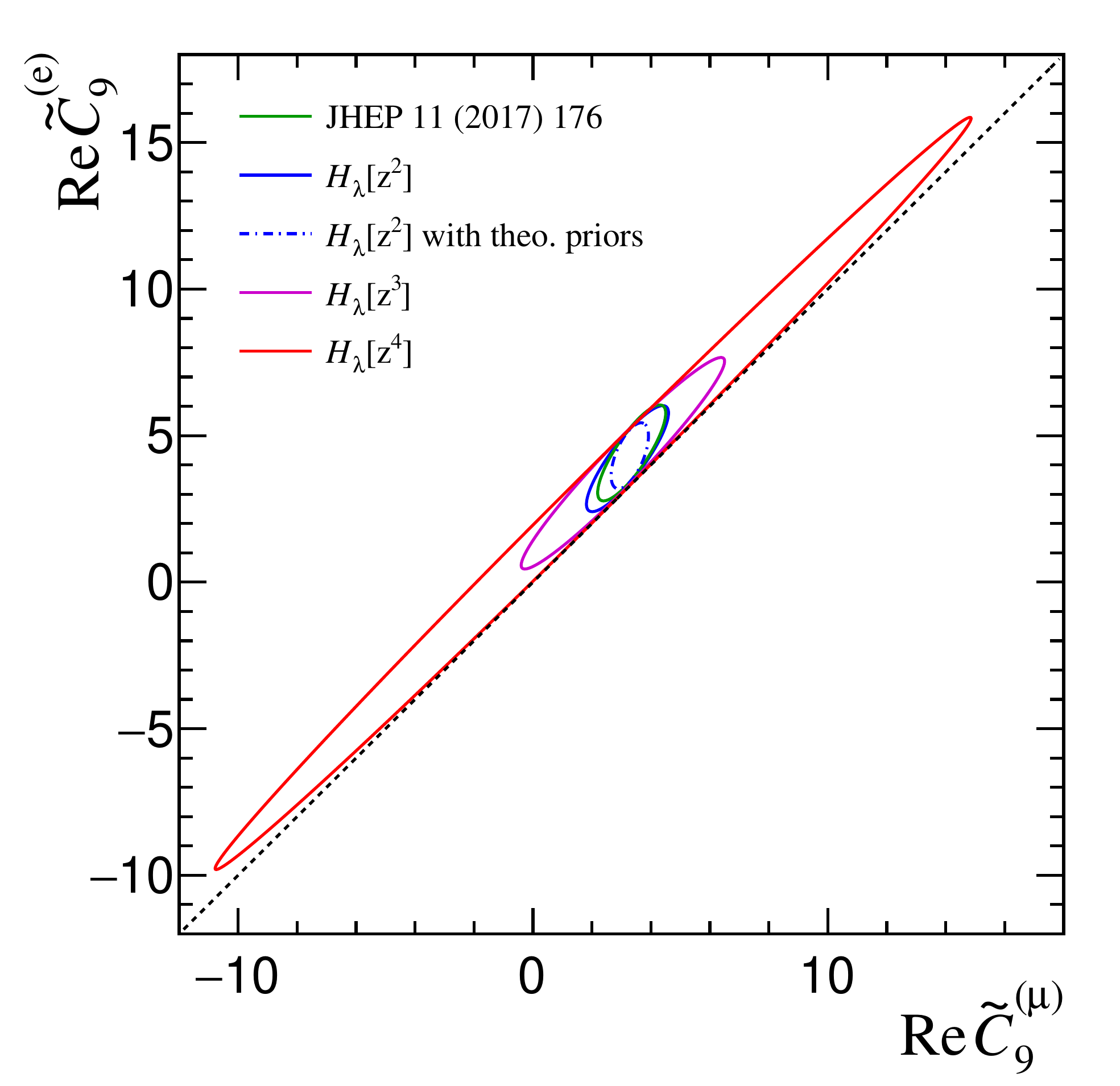} 
\caption{%
    Two-dimensional sensitivity scans for the pair of Wilson coefficients 
    $\widetilde{\mathcal{C}}_9^{(\mu)}$ and $\widetilde{\mathcal{C}}_9^{(e)}$ 
    for different nonlocal hadronic parametrisation models evaluated at \texttt{BMP}$_{\WC_9}$,  
    and with the expected statistics after \lhcb run II. 
    The contours correspond to 99\% confidence level statistical-only uncertainty bands and 
    the dotted black line indicates the LFU hypothesis.   
}
\label{fig:C9ellipse}
\end{figure}

Figure~\ref{fig:C9ellipse} shows the fit results for several alternative parametrisations 
of the nonlocal hadronic contribution for the \texttt{BMP}$_{\WC_9}$ hypothesis, 
with yields corresponding to LHCb run II scenario.  
We observe that the sensitivity to $\widetilde{\WC}_9^{(\mu,e)}$ is strongly dependent on 
the model assumption used for the nonlocal matrix elements. 
Nonetheless, it is noticeable that the high correlation of the  
$\widetilde{\mathcal{C}}_9^{(\mu)}$ and $\widetilde{\mathcal{C}}_9^{(e)}$ coefficients 
is sufficient to preserve the true underlying physics at any order of the series expansion $\mathcal{H}_\lambda[z^n]$
and without any parametric theoretical input,  
i.e., the two-dimensional pull estimator with respect to the LFU hypothesis is unbiased. 
We note that, as commonly stated in the literature (see e.g., recent review in Ref.~\cite{Capdevila:2017ert}),  
the determination of $\WC_{10}^{(\mu,e)}$ is insensitive to the lack of knowledge on the 
nonlocal hadronic effects. Nevertheless, its precision is still bounded to the uncertainties on the form factors, 
that are found to be the limiting factor by the end of run II. 
\begin{figure}[bth!]
\includegraphics[width=.4\textwidth]{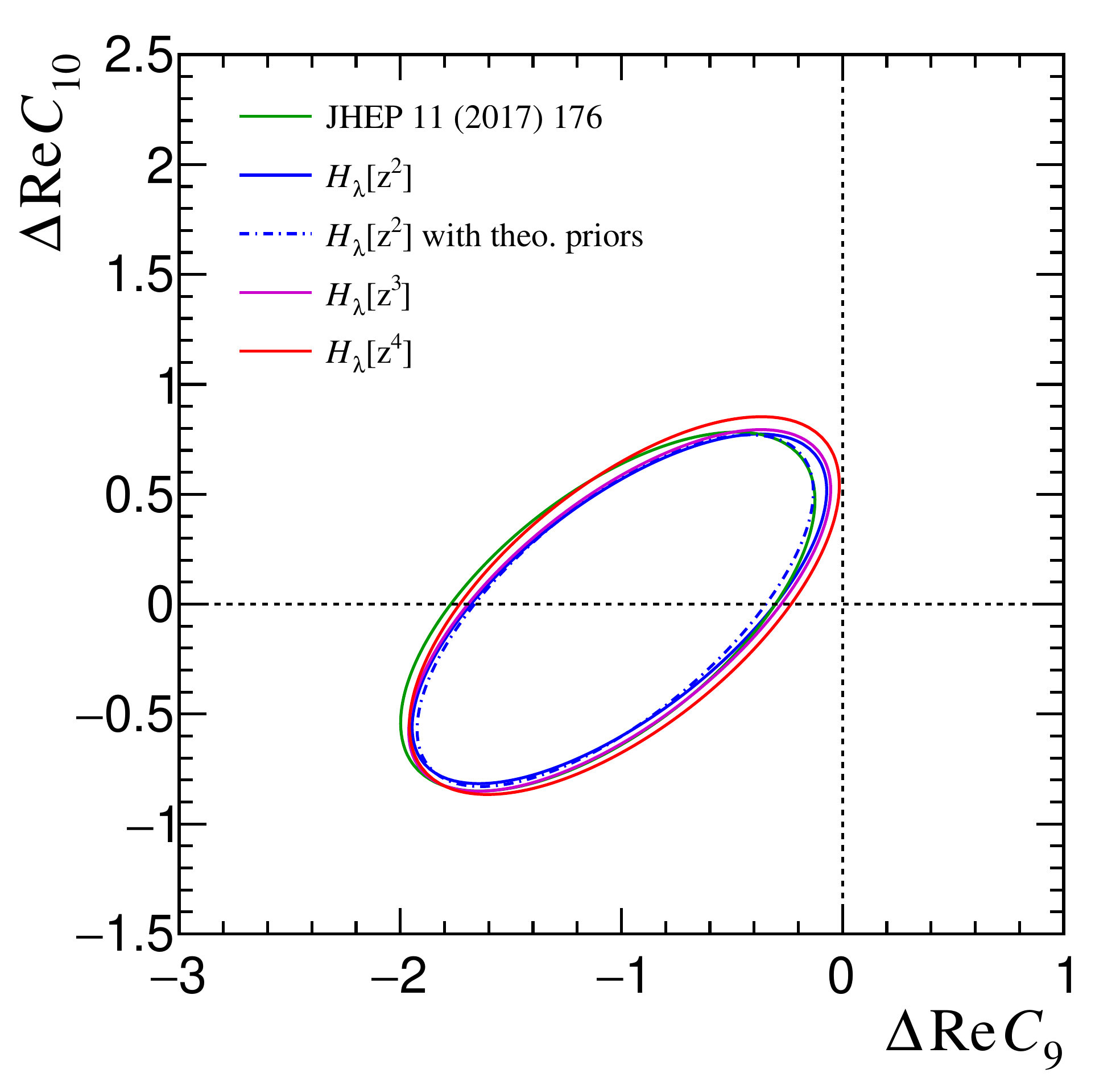}\\
\includegraphics[width=.4\textwidth]{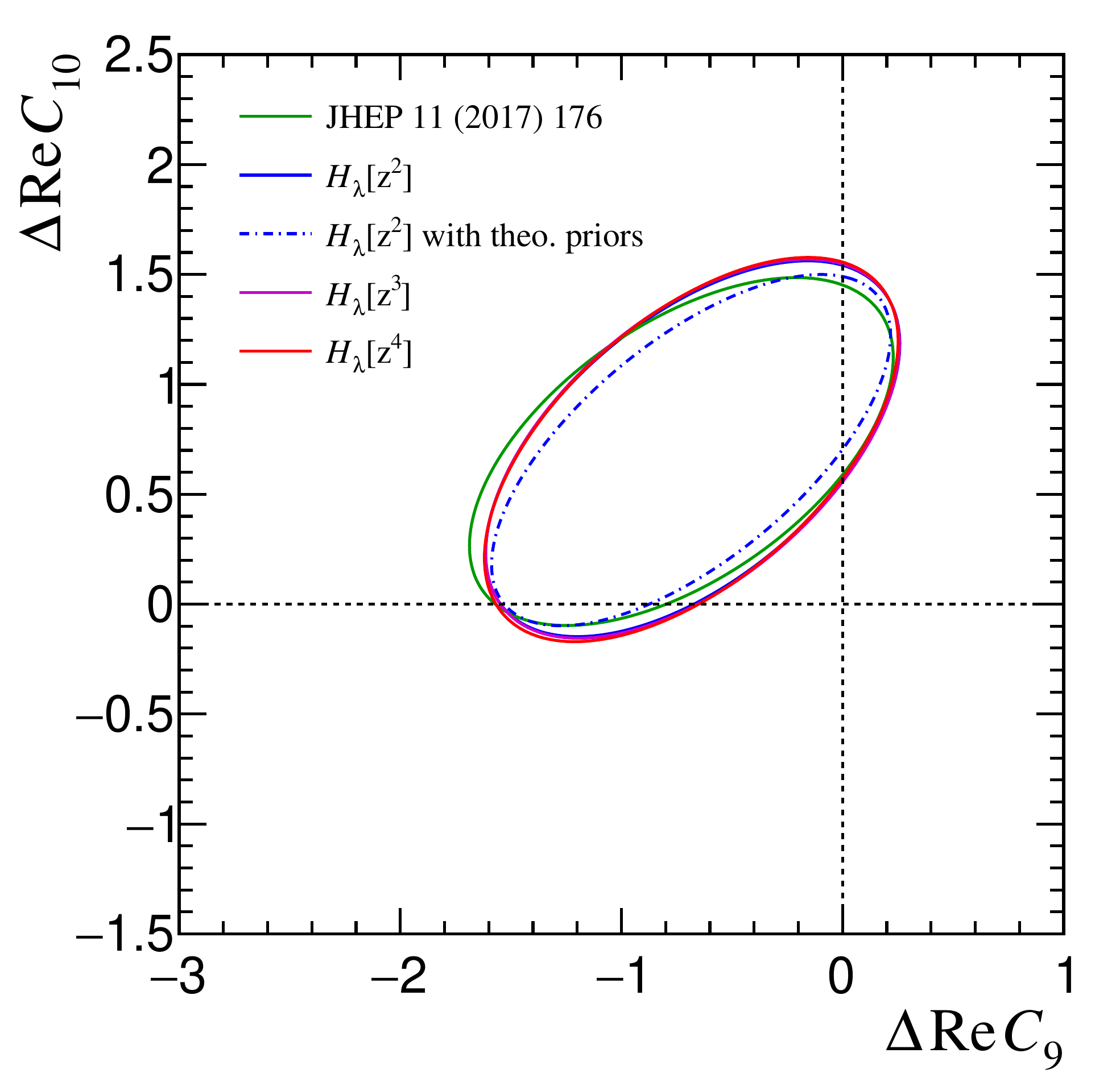} 
\caption{%
    Two-dimensional sensitivity scans for the proposed observables $\Delta\WC_9$ and $\Delta\WC_{10}$ 
    for different nonlocal hadronic parametrisation models 
    evaluated at (top) \texttt{BMP}$_{\WC_9}$ and (bottom) \texttt{BMP}$_{\WC_{9,10}}$, 
    and with the expected statistics after \lhcb run II. 
    The contours correspond to 99\% confidence level statistical-only uncertainty bands.
    \label{fig:DeltaC9C10}
}
\end{figure}

\begin{figure}[bth!]
\includegraphics[width=.4\textwidth]{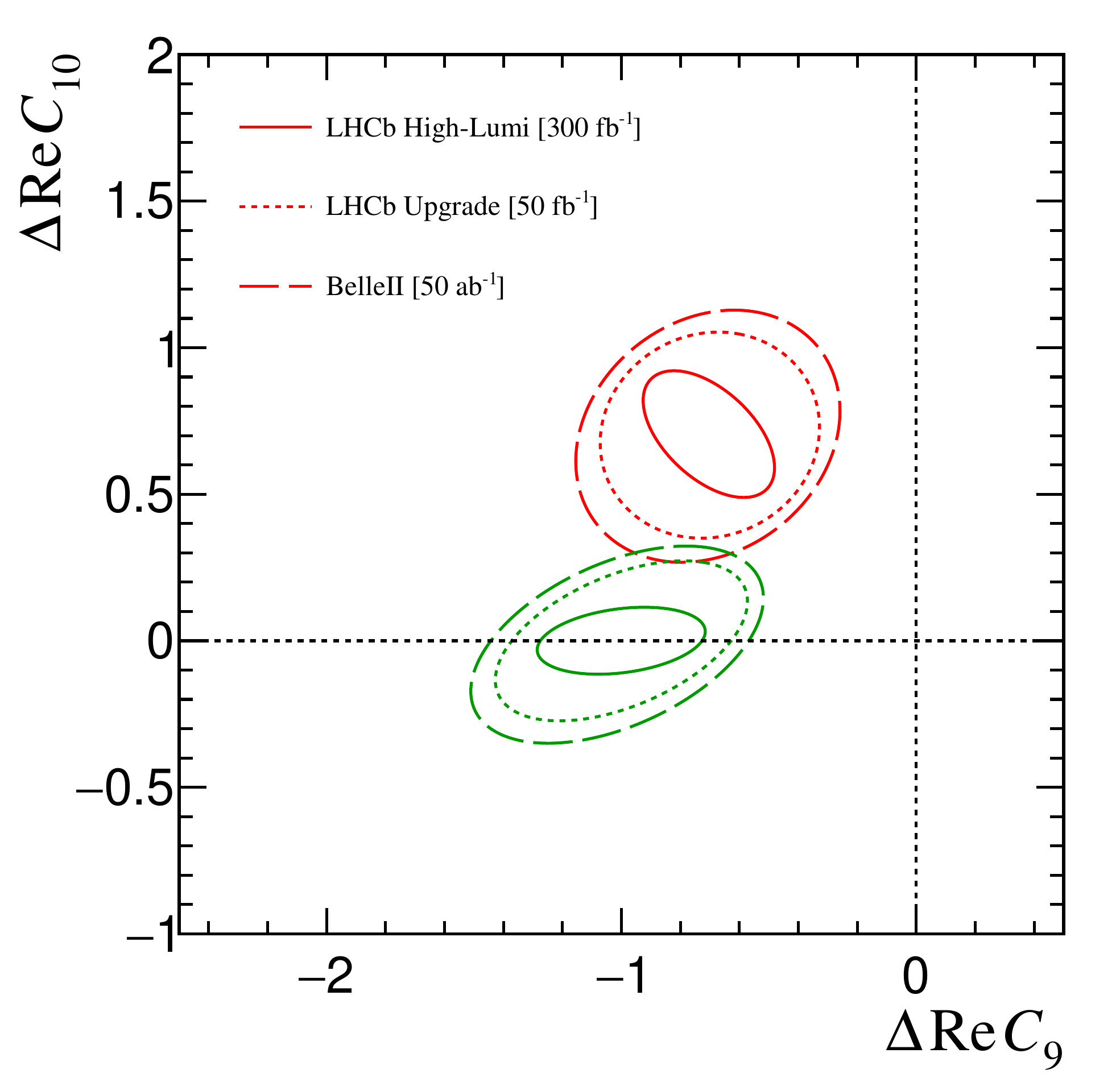}
\caption{%
    Two-dimensional sensitivity scans for the proposed observables $\Delta\WC_9$ and $\Delta\WC_{10}$ 
    for the two considered NP scenarios: (green) \texttt{BMP}$_{\WC_9}$ and (red) \texttt{BMP}$_{\WC_{9,10}}$.
    The contours correspond to 99\% confidence level statistical-only uncertainty bands expected for the (dashed) \belle II 50~ab$^{-1}$ 
    and \lhcb Upgrade (dotted) $50\,$fb$^{-1}$  and (solid) $\,300\,$fb$^{-1}$ statistics.
    \label{fig:DeltaC9C10_Upgrade}
}
\end{figure}

The sensitivity to the two benchmarklike NP scenarios using the proposed pseudo-observables $\Delta \WC_i$
is shown in Fig.~\ref{fig:DeltaC9C10}. 
We quantify the maximal expected significance with respect to the SM to be $4.6$ and $5.3\,\sigma$ for 
\texttt{BMP}$_{\WC_9}$ and \texttt{BMP}$_{\WC_{9,10}}$, respectively. 
Realistic experimental effects are necessary to determine the exact sensitivity achievable.
Nevertheless, these results suggest that a first observation (with a single measurement) of LFU breaking 
appears to be feasible with the expected recorded statistics by the end of LHCb run II. 
Furthermore, it is interesting to examine the prospects for confirming this evidence in the upcoming LHCb/Belle 
upgrades~\cite{Albrecht:2017odf}. 
Figure~\ref{fig:DeltaC9C10_Upgrade} summarises the two-dimensional statistical-only significances  
for the designed luminosities. 
Both LHCb Upgrade and Belle II experiments have comparable sensitivities (within $8.0-10\,\sigma$), 
while LHCb High-Lumi has an overwhelming significance.  
These unprecedented data sets will not only yield insights on this phenomena but also 
enable a deeper understanding of the nature of NP - 
insensitive to both local and nonlocal hadronic uncertainties.

Experimental resolution and detector acceptance/efficiency effects  
are not considered in this work, as these would require further information
from current (nonpublic) or planned \textit{B}-physics experiments.
Nevertheless, the precision on this measurement can remain unbiased 
either by parametrising these effects in the amplitude model and/or even
recomputing the angles or the $q^2$ variables constraining the $B$ invariant mass~\cite{Lionetto:2624938}.
Moreover, the differential decay width can receive additional complex amplitudes from signal-like backgrounds,  
e.g., $K\pi$ S-wave from a nonresonant decay and/or a scalar resonance~\cite{Becirevic:2012dp}. 
These contributions are determined to be small~\cite{Aaij:2015oid,Aaij:2016flj}, 
and in the proposed formalism they benefit from the same description between the muon 
and electron mode (see detailed discussion in Ref.~\cite{Chrzaszcz:2018yza}).
Therefore, such contribution does not dilute the expected sensitivity of the measurement.

Another important test to probe the stability of the model consists in 
analysing potential issues that can rise if the truncation 
$\mathcal{H}_\lambda[z^n]$ is not a good description of nature.
We proceed as follows: we generate ensembles with nonzero coefficients for 
$\mathcal{H}_\lambda[z^3]$ and $\mathcal{H}_\lambda[z^4]$, and we perform the fit 
with $\mathcal{H}_\lambda[z^2]$.
Despite the mismodelling of the nonlocal hadronic effects in the fit, we observe 
that the determination of $\Delta\WC_9$ and $\Delta\WC_{10}$ is always unbiased, 
thanks to the relative cancellation of all the shared parameters between the two channels.
It is worth mentioning that a hypothetical determination of the individual  
$\widetilde{\WC}_9^{(\mu,e)}$ and $\widetilde{\WC}_{10}^{(\mu,e)}$ WCs 
can also produce a shift in their central values
that mimics the behaviour of NP~\cite{Chrzaszcz:2018yza}.

\begin{figure}[t]
\includegraphics[width=.4\textwidth]{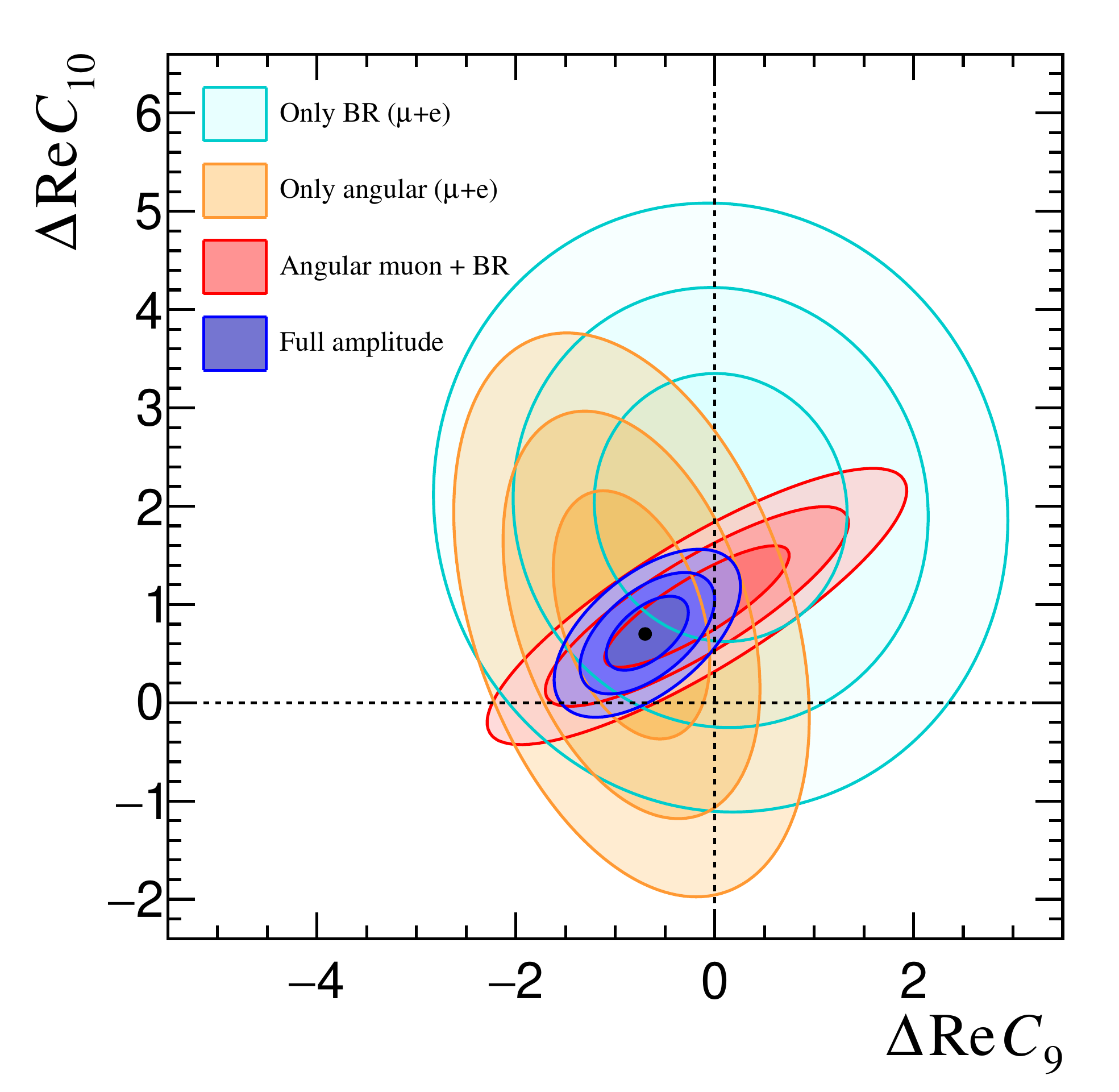} 
\caption{%
    Sensitivity to \texttt{BMP}$_{\WC_{9,10}}$ scenario for the expected statistics after the \lhcb run II.
    The relative contribution (68, 95, 99\% confidence level contours) of each step of the analysis is shown in different colours, 
    together with the result of full amplitude method proposed in this article. 
    \label{fig:allComponents}
}
\end{figure}

In conclusion, we propose a clean and model-independent method to combine
all the available information from $\bar{B}\to \bar{K}^*\ell^+\ell^-$ decays for a precise 
determination of LFU-breaking differences of WCs, i.e., $\Delta\WC_9$ 
and $\Delta\WC_{10}$.
This relies on a shared parametrisation of the local (form factors) and nonlocal ($\mathcal{H}_\lambda[z^n]$)
hadronic matrix elements between the muonic and electronic channels, 
that in turn enables the determination of the observables of interest
free from any theoretical uncertainty.  
In addition, this simultaneous analysis is more robust against 
experimental effects such
as mismodeling of the detector resolution, since most parameters are
effectively determined from the muon mode. This would be an important
benefit for LHCb where the electron resolution is significantly worse
than that of muons. 
Figure~\ref{fig:allComponents} illustrates the usefulness of the newly-proposed observables by combining 
the different information from angular analysis to branching ratio measurements. 
Due to the inclusiveness of the approach, the expected sensitivity surpasses any 
of the projections for the foreseen measurements of, e.g., $R_{K^{*}}$ or $P^{\prime}_{5}$ alone - given the benchmark points.
Therefore, this novel formalism can be the most immediate method to observe unambiguously 
NP in $\bar{B}\to \bar{K}^*\ell^+\ell^-$ decays.

A promising feature of this framework is the possibility to extend the analysis to 
include other decay channels involving flavour changing neutral currents. 
For instance, the charged decay $\bar{B}^+ \to \bar{K}^{*+} \ellell$ undergoes the same physics 
and is easily accessible at the $B$-factories, while other rare 
semi-leptonic decays such as $B^+ \to K^+ \ellell$ and $\Lambda_{b} \to \Lambda^{(*)} \ell^+\ell^-$ 
have a different phenomenology but access the same NP information in terms of WC description.
Thus, an unbinned global simultaneous fit to all data involving $b \to s \ell^+ \ell^-$ transitions
is a natural and appealing extension of this work. 
Moreover, the parameter space of the investigated WCs can also be broadened to 
incorporate direct measurement of the right-handed $\WC_i^{\prime}$ - 
currently weakly constrained by global fits~\cite{Capdevila:2017bsm,Altmannshofer:2017yso,Hurth:2017hxg}. 

\section*{Acknowledgments}

We acknowledge useful contributions from Michele Atzeni, Gino Isidori, Danny van Dyk and Patrick Owen.
This work is supported by the Swiss National Science Foundation (SNF) under Contracts No. 173104 and No. 174182.


\appendix

\section{Formalism}\label{app:formalism}

The nonlocal hadronic matrix elements $\mathcal{H}_\lambda(q^2)$ are investigated 
using two complementary parametrisations~\cite{Bobeth:2017vxj,Hurth:2017sqw}.

The nominal parametrisation~\cite{Bobeth:2017vxj} is obtained through the mapping
\begin{equation}
q^{2} \mapsto z(q^{2}) \equiv \frac{\sqrt{t_+ - q^2} - \sqrt{t_+ - t_0}}{\sqrt{t_+ - q^2} + \sqrt{t_+ - t_0}} \, ,
\end{equation}
where $t_+ = 4 M_D^2$ and $t_0 = t_+ - \sqrt{t_+ ( t_+ - M^2_{\psi(2S)})}$,
which leads to the functions $\mathcal{H}_\lambda(z)$ that are characterised 
by two singularities at $z_{J/\psi}$ and $z_{\psi(2S)}$.
These can be expressed as
\begin{equation}
\mathcal{H}_\lambda(z) = \frac{1 - z z^*_{J/\psi}}{z - z_{J/\psi}} \frac{1 - z z^*_{\psi(2S)}}{z - z_{\psi(2S)}}  \hat{\mathcal{H}}_\lambda(z)  \, ,
\end{equation}
where the functions $ \hat{\mathcal{H}}_\lambda(z)$ are analytical and can be Taylor-expanded around $z = 0$ as
\begin{equation}
\hat{\mathcal{H}}_\lambda(z)   =  \Big[ \sum^n_{k=0} \alpha_k^{(\lambda)} z^k \Big]  \mathcal{F}_\lambda(z)  \, .
\end{equation}
Several orders of the polynomials are studied in the text.
Note that any additional order $k$ introduces a complex parameter, $\alpha_k^{(\lambda)}$, for each of the polarisations $\lambda = \perp, \parallel, 0$.
These nuisance parameters can be either free floated in the fit (nominal configuration labelled as $\mathcal{H}_\lambda[z^2,...,z^4]$)
or Gaussian constrained to their SM prediction (labelled as $\mathcal{H}_\lambda[z^2]$ with theo. priors in the plots).

Finally, the nonlocal hadronic matrix elements are reparametrised following Ref.~\cite{Hurth:2017sqw}, in which 
these nonlocal hadronic contributions are included as multiplicative factors, leading to a reformulation 
of the amplitudes of Eq.~\ref{eq:amp_dep} as
\eqa{
  {\cal{A}}_{\lambda}^{(\ell)\,L,R} = {\cal{N}}_{\lambda}^{(\ell)}\ \bigg\{ &&
(C^{(\ell)}_9 \mp C^{(\ell)}_{10}) {\cal{F}}_{\lambda}(q^2)  \big[    1 + a_{\lambda} + b_{\lambda} \frac{q^{2}}{\mathrm{\scalebox{0.8}{6 GeV$^2$}}}  \big]  \nonumber   \\
&&  +\frac{2m_b M_B}{q^2} C^{(\ell)}_7 {\cal{F}}_{\lambda}^{T}(q^2) 
\bigg\}\,,
}
where $a_{\lambda}$ and $b_{\lambda}$ are complex coefficients Gaussian constrained around zero.

\section{Right-handed Wilson coefficients}\label{app:WC_primed}

An extension of the physics case of the proposed method
is to investigate the sensitivity to the chirality-flipped counterparts of the usual Wilson coefficients, 
i.e., $\WC^{\prime (\mu)}_9$ and $\WC^{\prime(\mu)}_{10}$.
Following the formalism discussed in this article, the primed WCs are examined by considering in addition to the 
\texttt{BMP}$_{\WC_{9,10}}$ three different modified NP scenarios for the muon only:
$\WC_{9,10}^{\prime(\mu)} = \WC^{\prime \rm{SM}}_{9,10} = 0$;
$\WC^{\prime (\mu)}_9 = \WC^{\prime (\mu)}_{10} = 0.3$; 
and $\WC^{\prime (\mu)}_9 = - \WC^{\prime (\mu)}_{10} =  0.3$.  
Notice that for the electron mode the $\WC_{9,10}^{\prime(e)}$ is set and fixed to the SM value $\WC^{\prime \rm{SM}}_{9,10} = 0$.

\begin{figure}[t]
\includegraphics[width=.4\textwidth]{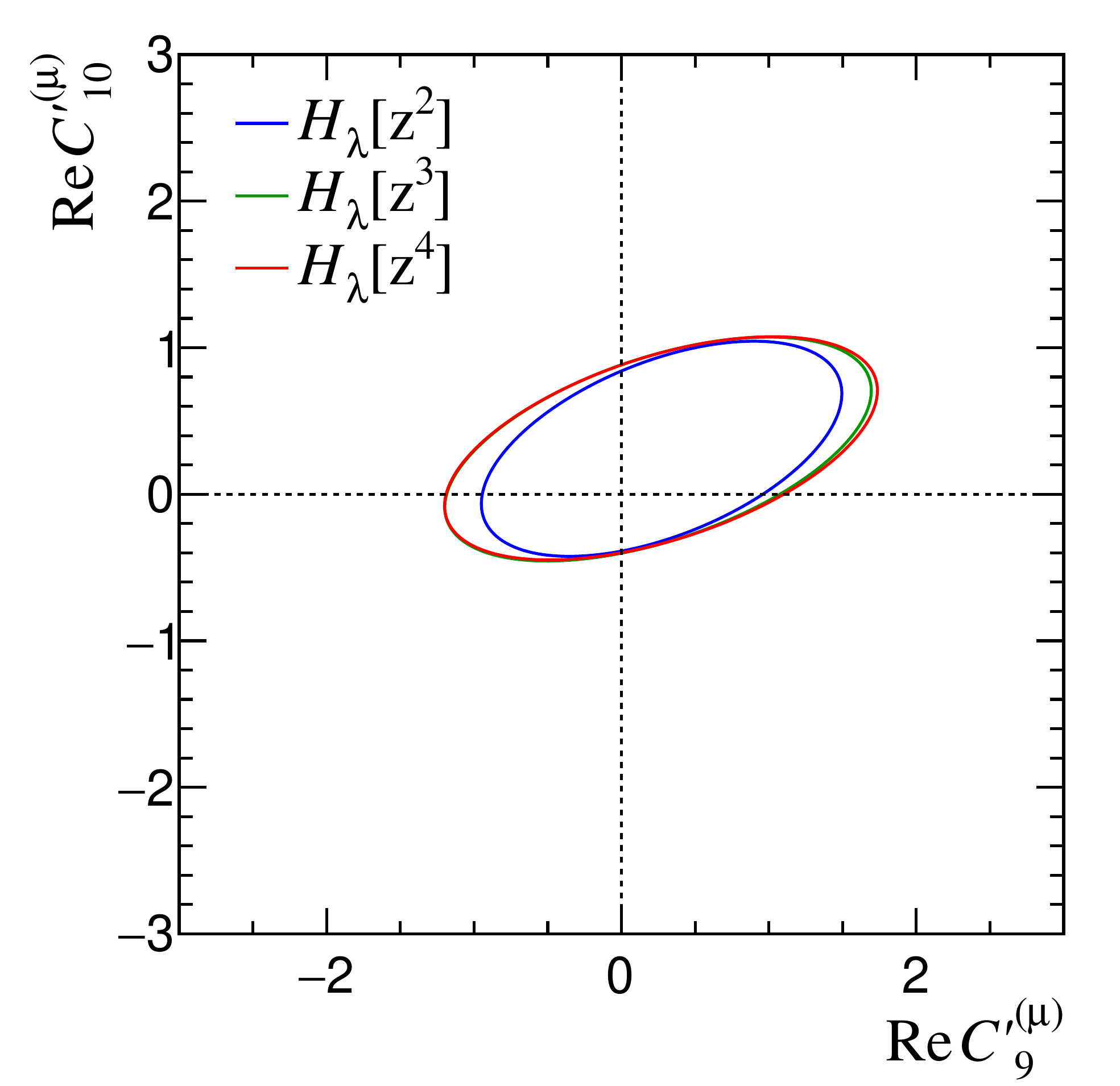} 
\caption{%
    Two-dimensional sensitivity scans for the pair of Wilson coefficients $\WC'^{(\mu)}_9$ and 
	$\WC'^{(\mu)}_{10}$ for different nonlocal hadronic parametrisation models for a NP scenario 
	with $\WC'^{(\mu)}_9 = \WC'^{(\mu)}_{10} = 0.3$.
    The contours correspond to 99\% confidence level statistical-only uncertainty bands evaluated with
	the expected statistics after \lhcb run II.
    \label{fig:Cp_Hz}
}
\end{figure}

\begin{figure}[t]
\includegraphics[width=.4\textwidth]{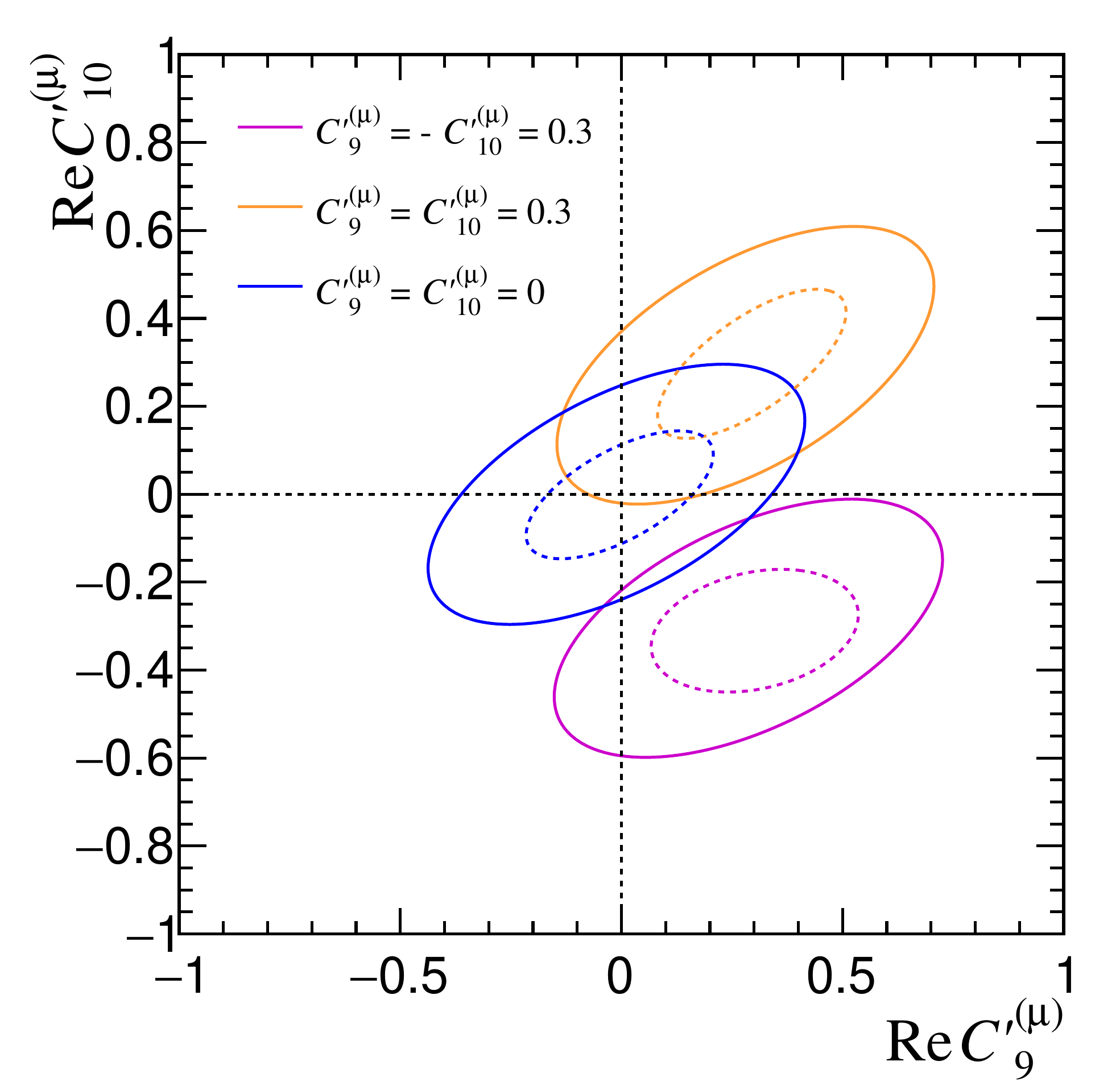} 
\caption{%
    Two-dimensional sensitivity scans for the pair of Wilson coefficients $\WC'^{(\mu)}_9$ and $\WC'^{(\mu)}_{10}$
    for three NP scenarios: (blue) $\WC'^{(\mu)}_9 = \WC'^{(\mu)}_{10} = 0$, (orange) $\WC'^{(\mu)}_9 = \WC'^{(\mu)}_{10} = 0.3$ 
    and (magenta) $\WC'^{(\mu)}_9 = - \WC'^{(\mu)}_{10} = 0.3$.
    The contours correspond to 99\% confidence level statistical-only uncertainty bands expected for the LHCb Upgrade (dotted)  
    $50\,$fb$^{-1}$ and (solid) $\,300\,$fb$^{-1}$ statistics.
    \label{fig:Cp}
}
\end{figure}

Figure~\ref{fig:Cp_Hz} shows the fit results for different order of the analytic expansion for 
the nonlocal hadronic contribution 
for a NP scenario with $\WC'^{(\mu)}_9 = \WC'^{(\mu)}_{10} = 0.3$, 
and yields corresponding to the \lhcb run II expected statistics. 
The dependency on the determination of $\WC'^{(\mu)}_9$ and $\WC'^{(\mu)}_{10}$ on the 
order of the expansion clearly saturates after $\mathcal{H}_\lambda[z^3]$ and allows a measurement
of the primed Wilson coefficients for the muon decay channel $B^{0} \to K^{*0} \mumu$ independent 
on the theoretical hadronic uncertainty.
Figure~\ref{fig:Cp} shows the prospects for the sensitivity to the $\WC'^{(\mu)}_9$ and 
$\WC'^{(\mu)}_{10}$ Wilson coefficients corresponding to the expected statistics at
the LHCb upgrade with $50$ and $300\,$fb$^{-1}$.
Note that only with the full capability of the LHCb experiment it is possible 
to start disentangling the different NP hypotheses. 

\FloatBarrier
\bibliography{references}

\end{document}